\documentclass[submission, Proceedings]{SciPost}

\begin{document}

\begin{center}{\Large \textbf{
Development of the detectors for the DeeMe experiment
}}\end{center}

% TODO: write the author list here. Use initials + surname format.
% Separate subsequent authors by a comma, omit comma at the end of the list.
% Mark the corresponding author with a superscript *.
\begin{center}
N. Teshima\textsuperscript{1*} on behalf of the DeeMe Collaboration
\end{center}

\begin{center}
{\bf 1} Osaka City University, Osaka, Japan
\\
* teshima@ocupc1.hep.osaka-cu.ac.jp
\end{center}

\begin{center}
\today
\end{center}

\definecolor{palegray}{gray}{0.95}
\begin{center}
\colorbox{palegray}{
  \begin{tabular}{rr}
  \begin{minipage}{0.05\textwidth}
    \includegraphics[width=8mm]{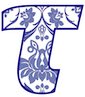}
  \end{minipage}
  &
  \begin{minipage}{0.82\textwidth}
    \begin{center}
    {\it Proceedings for the 15th International Workshop on Tau Lepton Physics,}\\
    {\it Amsterdam, The Netherlands, 24-28 September 2018} \\
    \href{https://scipost.org/SciPostPhysProc.1}{\small \sf scipost.org/SciPostPhysProc.Tau2018}\\
    \end{center}
  \end{minipage}
\end{tabular}
}
\end{center}

% For convenience during refereeing: line numbers
%\linenumbers

\section*{Abstract}
{\bf
% TODO: write your abstract here.
%The DeeMe experiment is planned to search for $\mu$-$e$ conversion in the nuclear field at the Materials and Life Science Experimental Facility in J-PARC. We aim to measure the process with a single event sensitivity of $10^{−14}$, which is one order of magnitude better than those of previous experiments. \\
%\indent
In the DeeMe experiment, approximately $70\ \mathrm{GHz/mm^{2}}$ prompt-charged particles will hit the multi-wire proportional chambers (MWPCs) between signal read-out periods. To avoid the space charge effect, we developed fast HV-switching MWPCs in order to control the gas gain dynamically. Four such MWPCs have been manufactured. The circuit of readout amplifiers is slightly modified to further improve the efficiency of the detector, and we also started to investigate other possible choices of the gas mixture. In this article, the development of the detectors and results of performance tests will be presented.
%The abstract is in boldface, and should fit in 8 lines.
%It should be written in a clear and accessible style, emphasizing the context, the problem(s) studied, the methods used, the results obtained, the conclusions reached, and the outlook. You can add a table contents, recommended if your paper is more than 6 pages long.
}

%\vspace{10pt}
%\noindent\rule{\textwidth}{1pt}
%\tableofcontents\thispagestyle{fancy}
%\noindent\rule{\textwidth}{1pt}
%\vspace{10pt}

\vspace{-20pt}
\section{Introduction}
\label{sec:intro}
% TODO: write your article here.
%
\vspace{-20pt}
\begin{center}
  \begin{figure}[b]
    \begin{tabular}{ccc}
      \begin{minipage}[t]{0.28\hsize}
        \centering
        \includegraphics[width=4.7cm,keepaspectratio,clip]{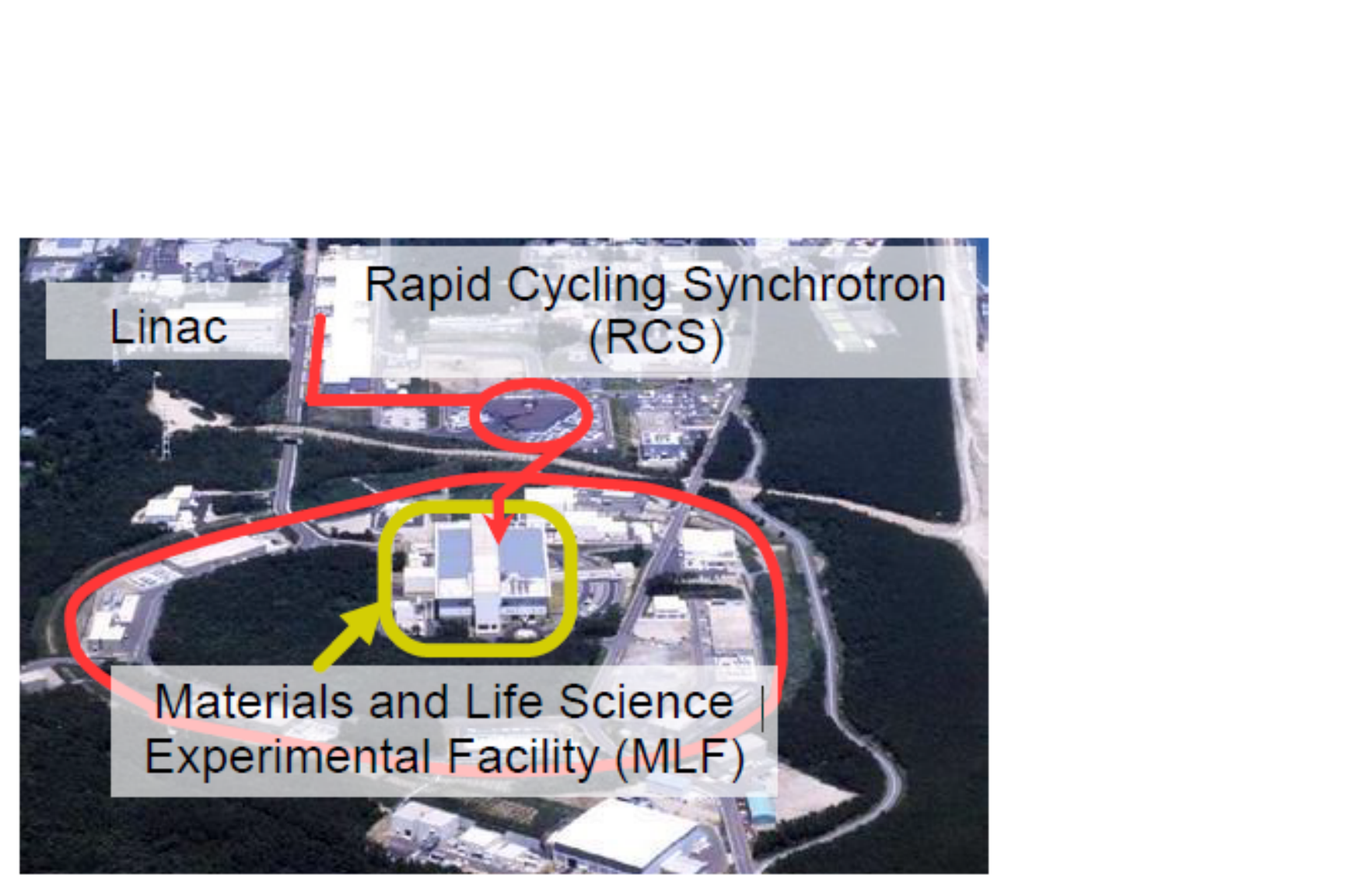}
        \caption{Photograph of the accelerators and Materials and Life Science Experimental Facility (MLF) at J-PARC.}
        \label{fig:jparcmlf}
      \end{minipage} &
      \begin{minipage}[t]{0.36\hsize}
        \centering
        \includegraphics[width=5.8cm,keepaspectratio,clip]{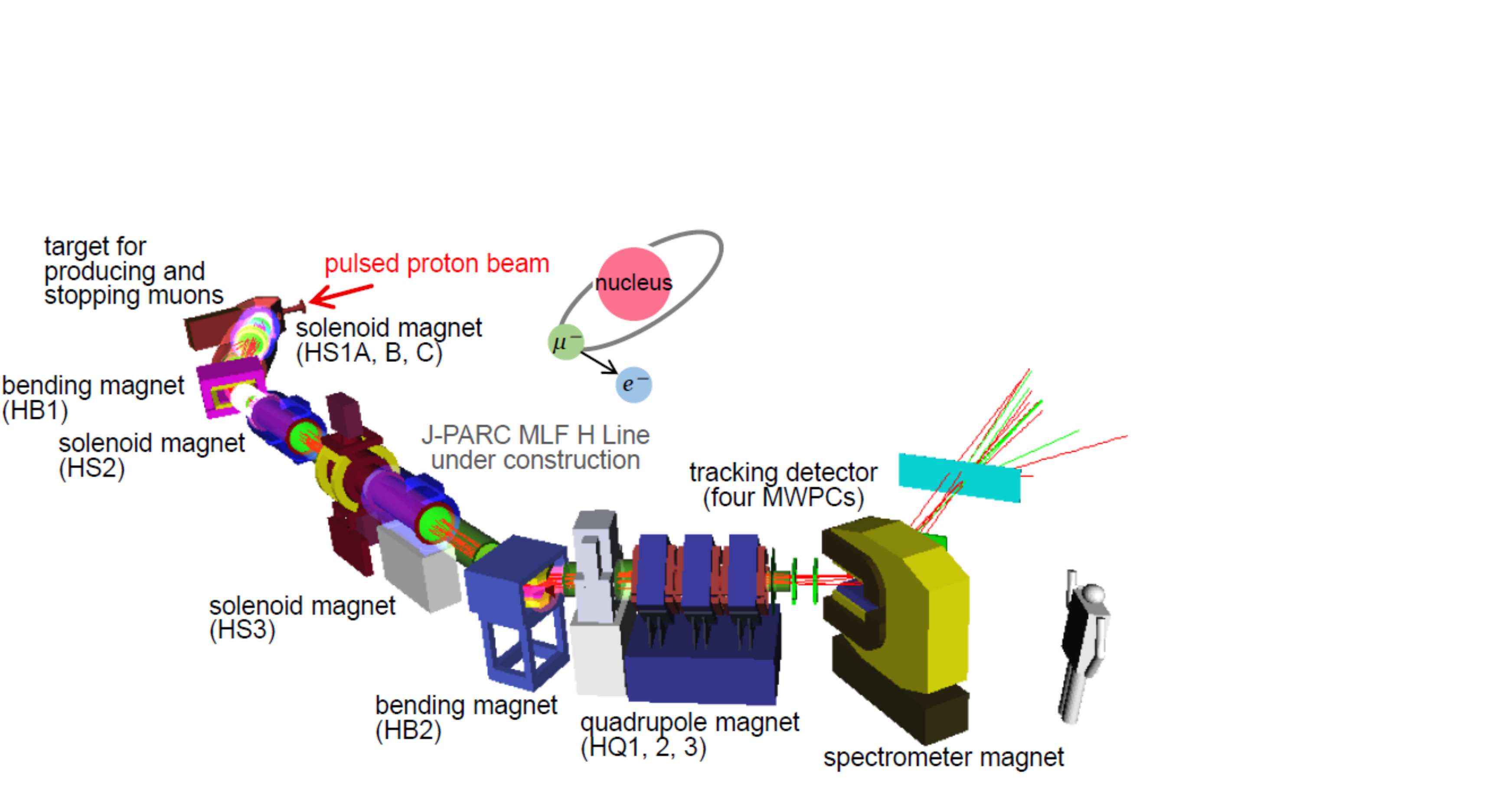}
        \caption{Simulated secondary beamline, H Line, in J-PARC MLF.}
        \label{fig:hlinesimu}
      \end{minipage} &
      \begin{minipage}[t]{0.28\hsize}
        \centering
        \includegraphics[width=4cm,keepaspectratio,clip]{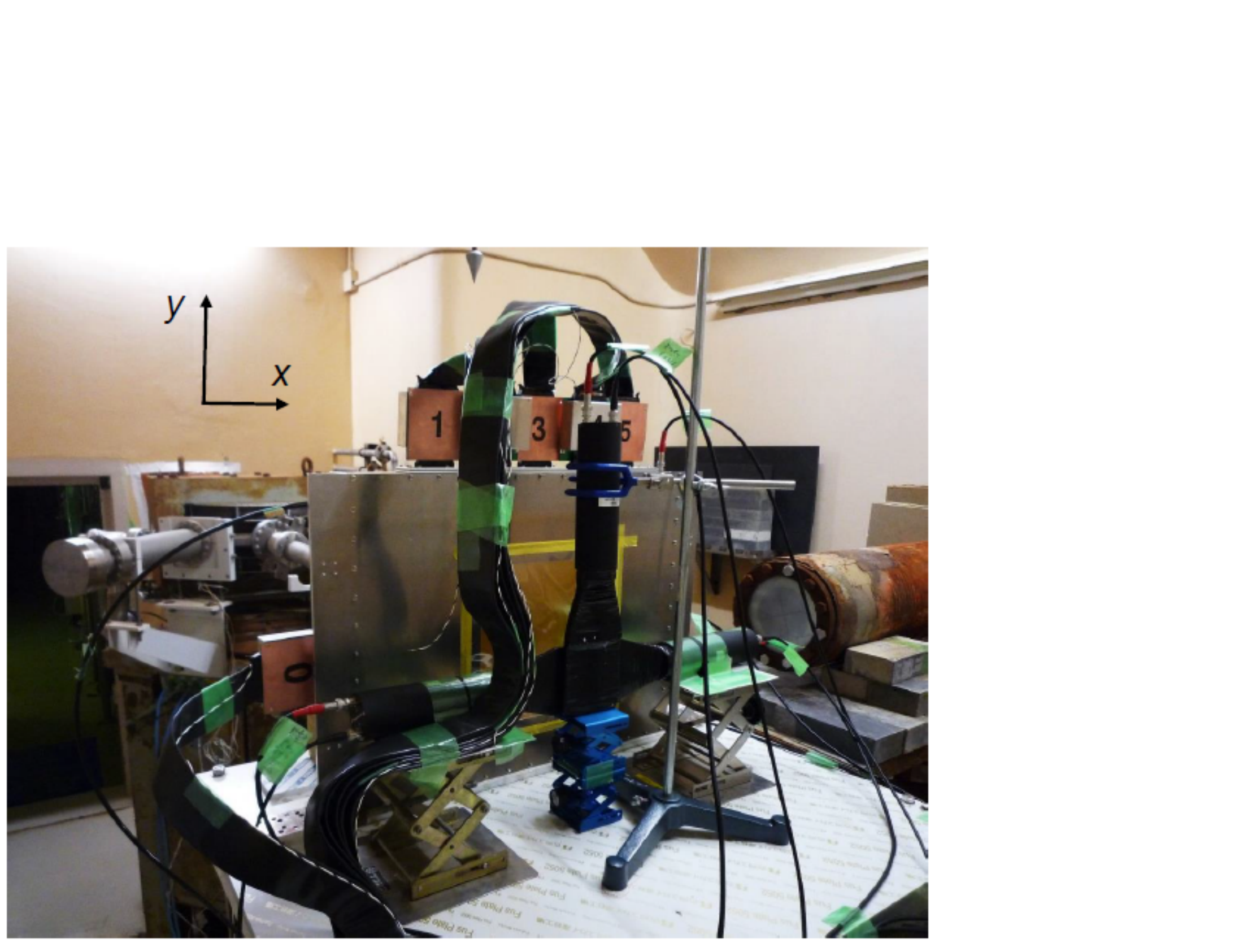}
        \caption{Photograph of one of the four multi-wire proportional chambers (MWPCs).}
        \label{fig:mwpcphoto}
      \end{minipage}      
    \end{tabular}
  \end{figure}
\end{center}
The DeeMe experiment is planned to search for muon-to-electron ($\mu$-$e$) conversion in the nuclear field at J-PARC Materials and Life Science Experimental Facility (MLF) H Line (see Fig. \ref{fig:jparcmlf} and \ref{fig:hlinesimu}). Our goal is to reach a single event sensitivity of $< 1\times10^{-13}$ for a graphite target or $< 2\times10^{-14}$ for a silicon carbide target with operating the Rapid Cycling Synchrotron (RCS) at a power of $1\ \mathrm{MW}$ for $2\times10^{7}\ \mathrm{sec/year}$. This will improve the sensitivity by one or two orders of magnitude than those achieved so far. \\
\vspace{-20pt}
\clearpage
\begin{center}
  \begin{figure}[t]
    \begin{tabular}{cc}
      \begin{minipage}[t]{0.4\hsize}
        \centering
        \includegraphics[width=6cm,keepaspectratio,clip]{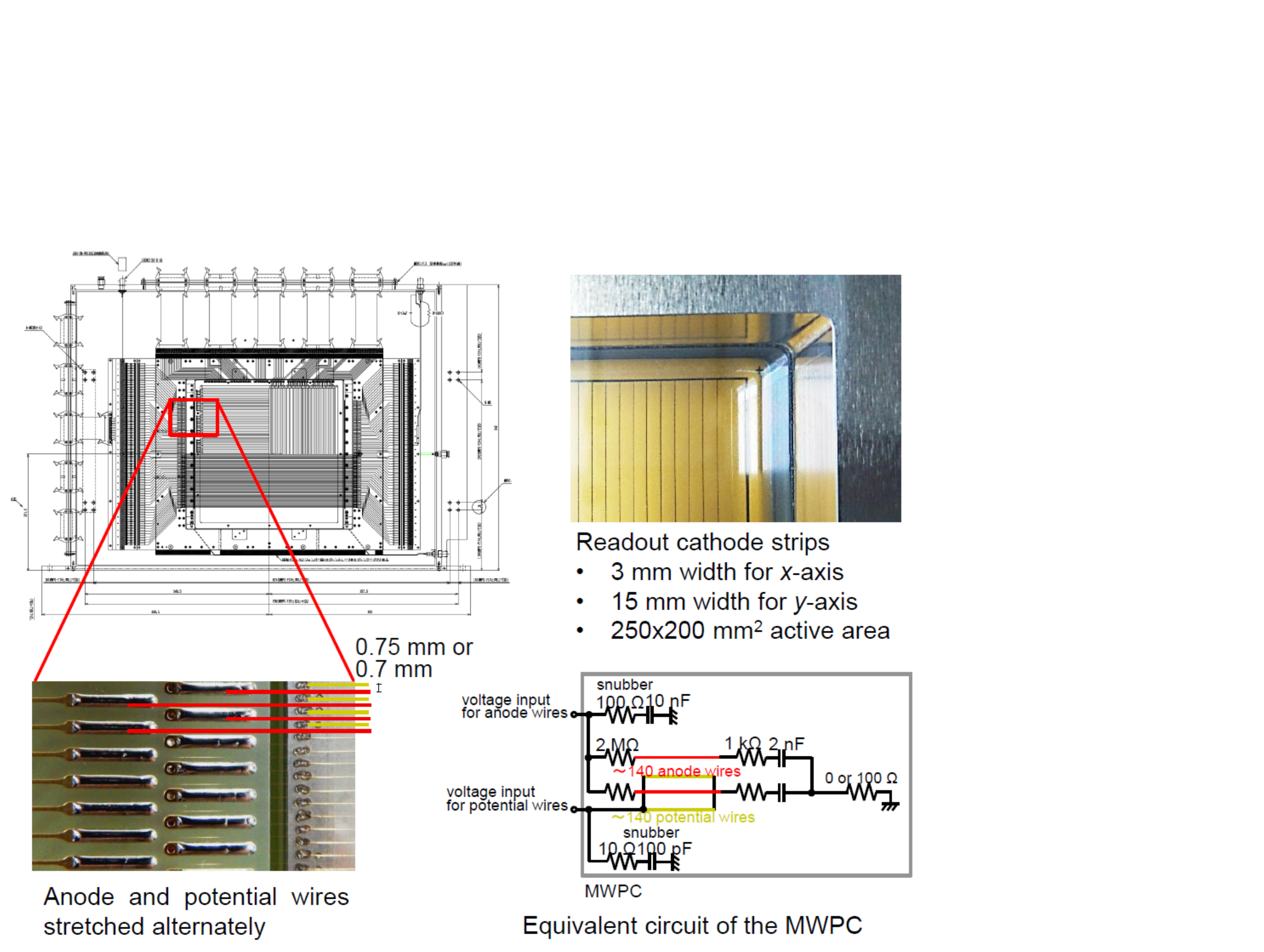}
        \caption{Structure of the MWPC.}
        \label{fig:mwpcstructure}
      \end{minipage} &
      \begin{minipage}[t]{0.55\hsize}
        \centering
        \includegraphics[width=8.2cm,keepaspectratio,clip]{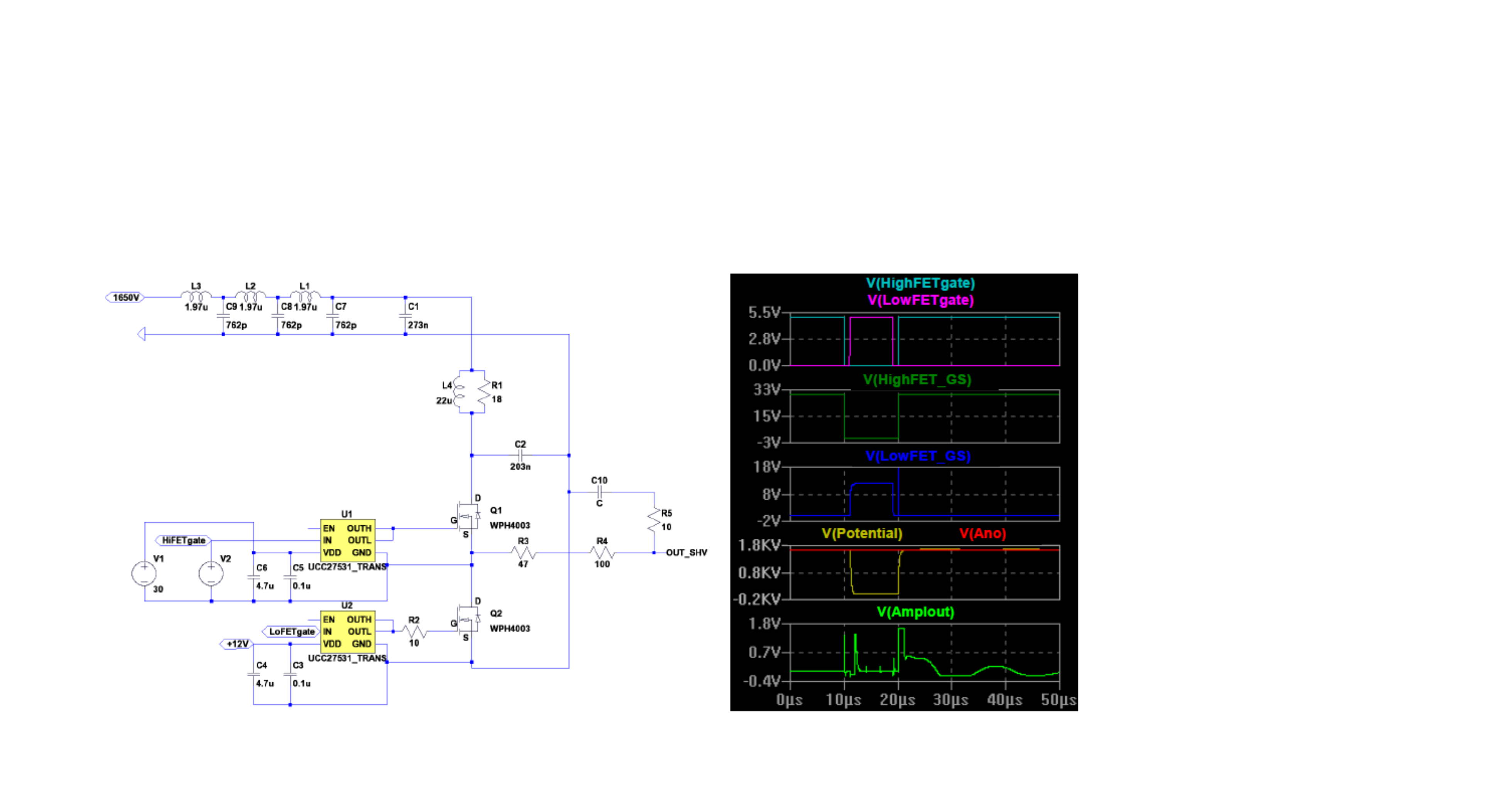}
        \caption{The circuit of high-voltage switching module and its simulation result.}
        \label{fig:hvswitchingcircuit}
      \end{minipage}
    \end{tabular}
  \end{figure}
\end{center}
\vspace{-40pt}
\mbox{}\\
\indent
$\mu$-$e$ conversion is coherent neutrino-less conversion of a muon into an electron in the nuclear field $\mu$N$\rightarrow e$N. It is one of charged lepton flavor violating (cLFV) processes, which are forbidden in the Standard Model (SM). Some theoretical models beyond the SM however predict observable branching fractions. An observation of cLFVs means the existence of new physics. The energy of signal electron is approximately $105\ \mathrm{MeV}$, which the mass of muon is converted to. \\
\indent
In the DeeMe experiment, a combined production and stopping target will be used to produce muonic atoms, and that electrons and other particles are transported to a spectrometer. The momenta of charged particles are measured by it, which consists of a magnet and four multi-wire proportional chambers (MWPCs). One of the four MWPCs is shown in Fig. \ref{fig:mwpcphoto}.
\vspace{-5pt}
\section{Development of the Detectors}
In the experiment, approximately $70\ \mathrm{GHz/mm^{2}}$ or $2\times10^{8}\ \mathrm{charged\ particles/pulse}$ of prompt burst will hit the MWPCs between signal read-out periods. To avoid the space charge effect, we developed fast high-voltage-switching MWPCs in order to control the gas gain dynamically.
\subsection{Devices}
Figure \ref{fig:mwpcstructure} shows the structure of the MWPC. The MWPC is used with cathode readout. It has anode wires and potential wires stretched alternately. A DC high voltage is applied to the anode wires. To the potential wires, $0\ \mathrm{V}$ is applied in a time window of a few microseconds in which we search for a signal of the $\mu$-$e$ conversion, or a voltage as high as the voltage on the anode wires is applied to reduce the gas gain to the order of 1 \cite{1}.\\
\indent
We put a high voltage switching module between HV supply and the potential wires. Figure \ref{fig:hvswitchingcircuit} shows the circuit of the module and its simulation result. It has two MOSFETs for outputting high voltage or $0\ \mathrm{V}$.
\subsection{Current Status}
\begin{center}
  \begin{figure}[t]
    \begin{tabular}{cc}
      \begin{minipage}[t]{0.4\hsize}
        \centering
        \includegraphics[width=6.3cm,keepaspectratio,clip]{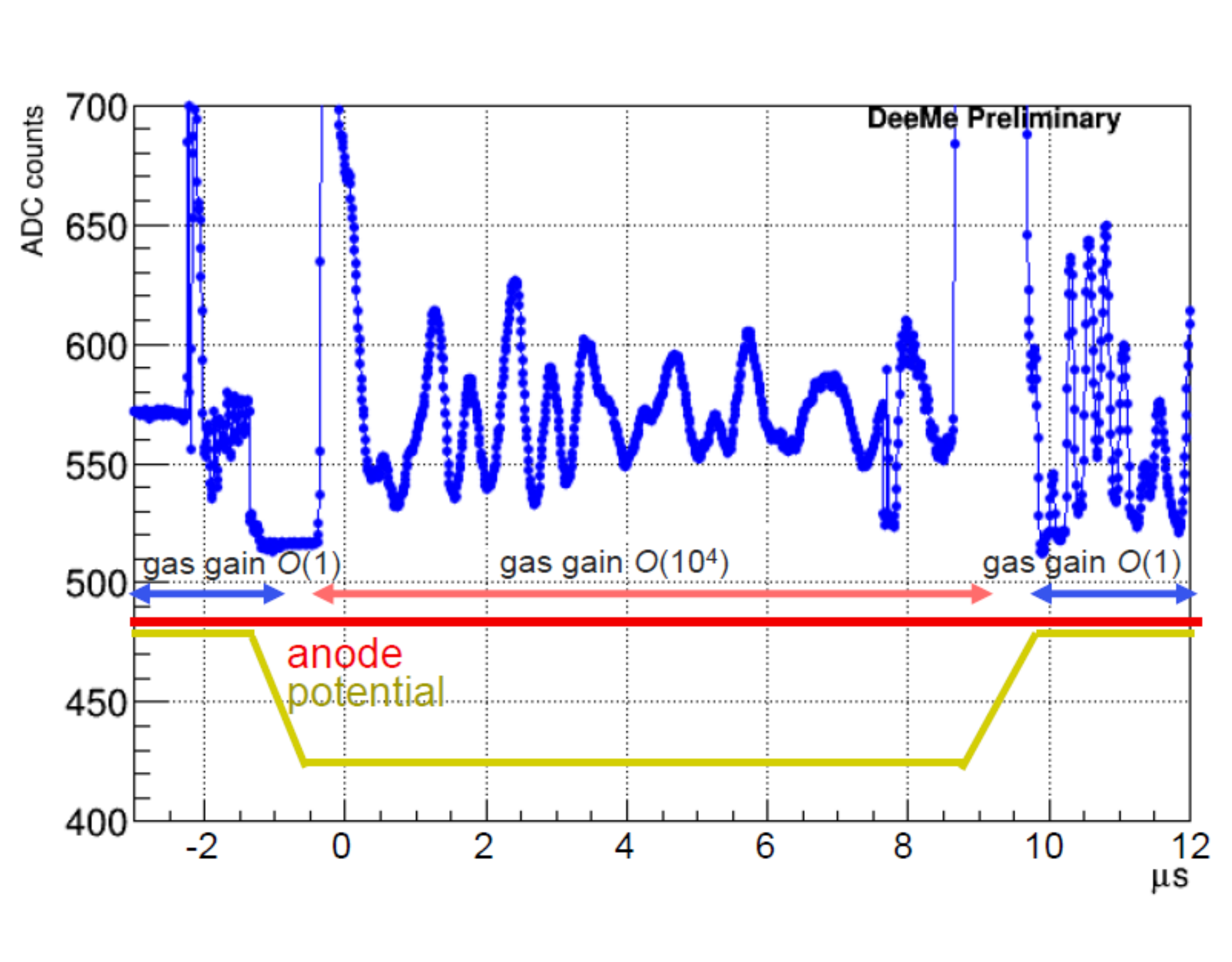}
        \caption{Output waveform of a cathode strip signal amplified through an amplifier.}
        \label{fig:wf}
      \end{minipage} &
      \begin{minipage}[t]{0.55\hsize}
        \centering
        \includegraphics[width=8.2cm,keepaspectratio,clip]{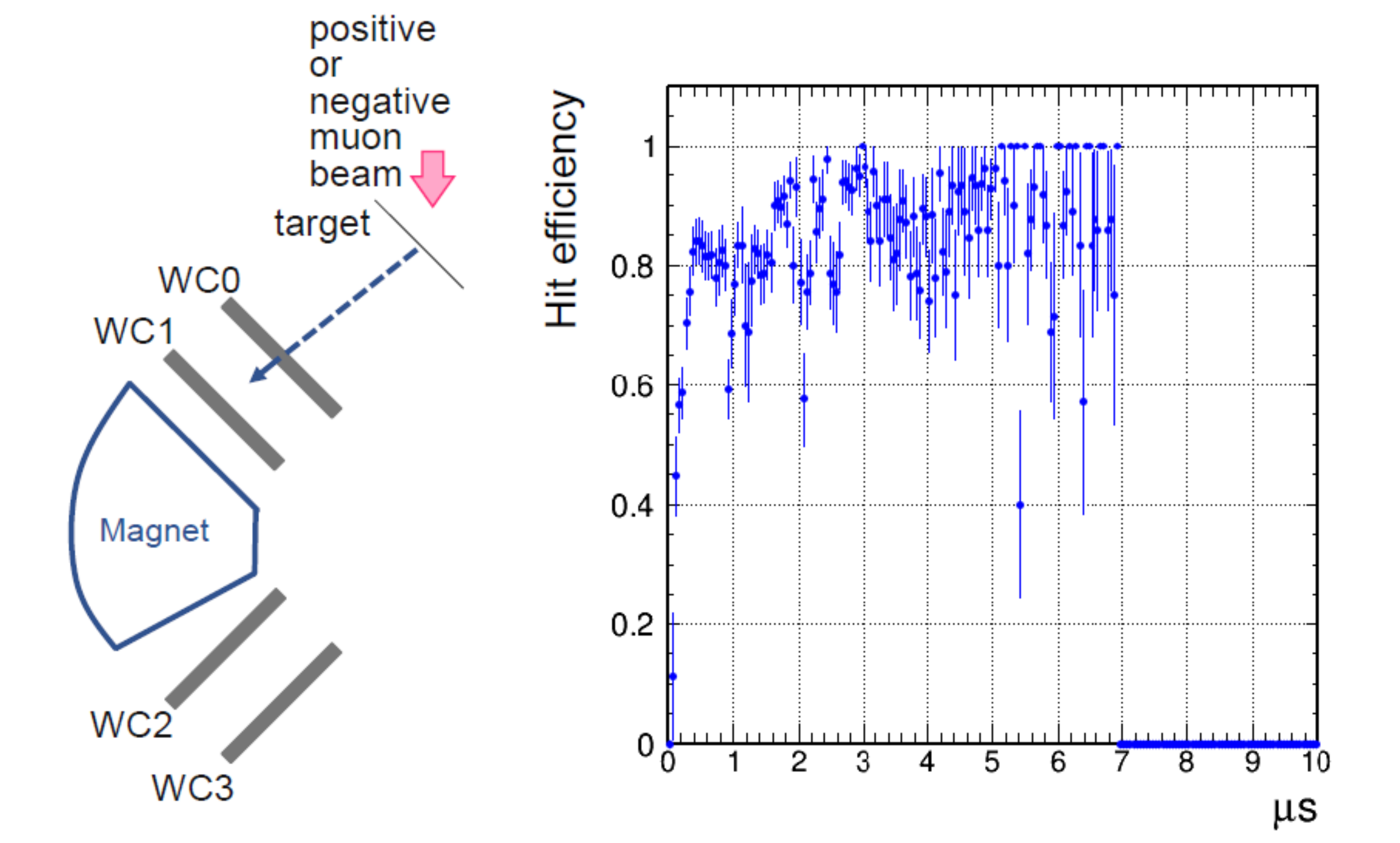}
        \caption{Setup (left) and hit efficiency of the second MWPC from the upstream (right) in the experiment at the J-PARC MLF D2 Area.}
        \label{fig:efficibefore}
      \end{minipage}
    \end{tabular}
  \end{figure}
\end{center}

For the gas mixture of argon (35\%) ethane (65\%) and applying $1630\ \mathrm{V}$ to the MWPC, a waveform obtained is shown in Fig. \ref{fig:wf}. In the period around $-1(+9)\ \mu\mathrm{s}$ the voltage on the potential wires is decreasing (increasing), resulting in negative (positive) saturation on the cathode strip readout. In between, the voltage on the potential wires is $0\ \mathrm{V}$, and the detector works with a gas gain of approximately $4.5\times10^{4}$, but there is an oscillation. The form of noise is however constant so that we can find a signal by subtracting the noise waveform. \\
\indent
In this operation, we conducted experiments at the J-PARC MLF D2 Area in March (three days) and June (five days), 2017 (see Fig. \ref{fig:efficibefore}). The purpose is to measure momenta of electrons from muon Decay in Orbit $\mu^{-} \rightarrow e^{-} \nu_{\mu} \overline{\nu_{e}}$ (DIO), one of the main background in the DeeMe experiment, about $50\ \mathrm{MeV/}c$. The hit efficiencies of x-axis (horizontal direction) readout of the MWPCs are analyzed to be $\simeq 90\%$ (WC0 and WC1 with $0.75\ \mathrm{mm}$ wire spacing and $1630\ \mathrm{V}$ applied) and $\simeq 60\%$ (WC2 and WC3 with $0.7\ \mathrm{mm}$ and $1600\ \mathrm{V}$ applied). Figure \ref{fig:efficibefore} (right) illustrates the efficiency of the second MWPC as a function of time. It fluctuates in time as the shape of output waveform oscillates. \\
\indent
To avoid loss of efficiency from negative saturation, we tried two things: (1) changing the filling gas and applying lower voltage and (2) increasing the dynamic range of the readout amplifiers.
\vspace{-20pt}
\subsubsection{Gas Mixture Study}
It is simulated that we can lower the voltage to $1510\ \mathrm{V}$ if we change the gas mixture into argon (80\%) isobutane (20\%) for the MWPC with a wire spacing of $0.75\ \mathrm{mm}$. \\
\indent
The stability of the MWPC depends on the tolerance to discharge between two kinds of wires. To check it, two wires were put in a small chamber and we investigated what voltage discharge occurs at \cite{2}. It was found to be at $\simeq 1950\ \mathrm{V}$. That means we have $\simeq 400\ \mathrm{V}$ margin for discharge when we choose the applied voltage $1510\ \mathrm{V}$.
\subsubsection{Amplifier Improvement}
Radeka-type two-stage amplifier is adopted at present. One stage consists of a common base and two emitter followers, and the amplifier has two stages \cite{3}. By changing the values of resistors of the second stage, we increased negative range of the amplifier from $\simeq 120\ \mathrm{mV}$ to $\simeq 280\ \mathrm{mV}$.
\clearpage
\section{Results of the Latest Beam Test}
\begin{center}
  \begin{figure}[t]
    \begin{tabular}{cc}
      \begin{minipage}[t]{0.55\hsize}
        \centering
        \includegraphics[width=9cm,keepaspectratio,clip]{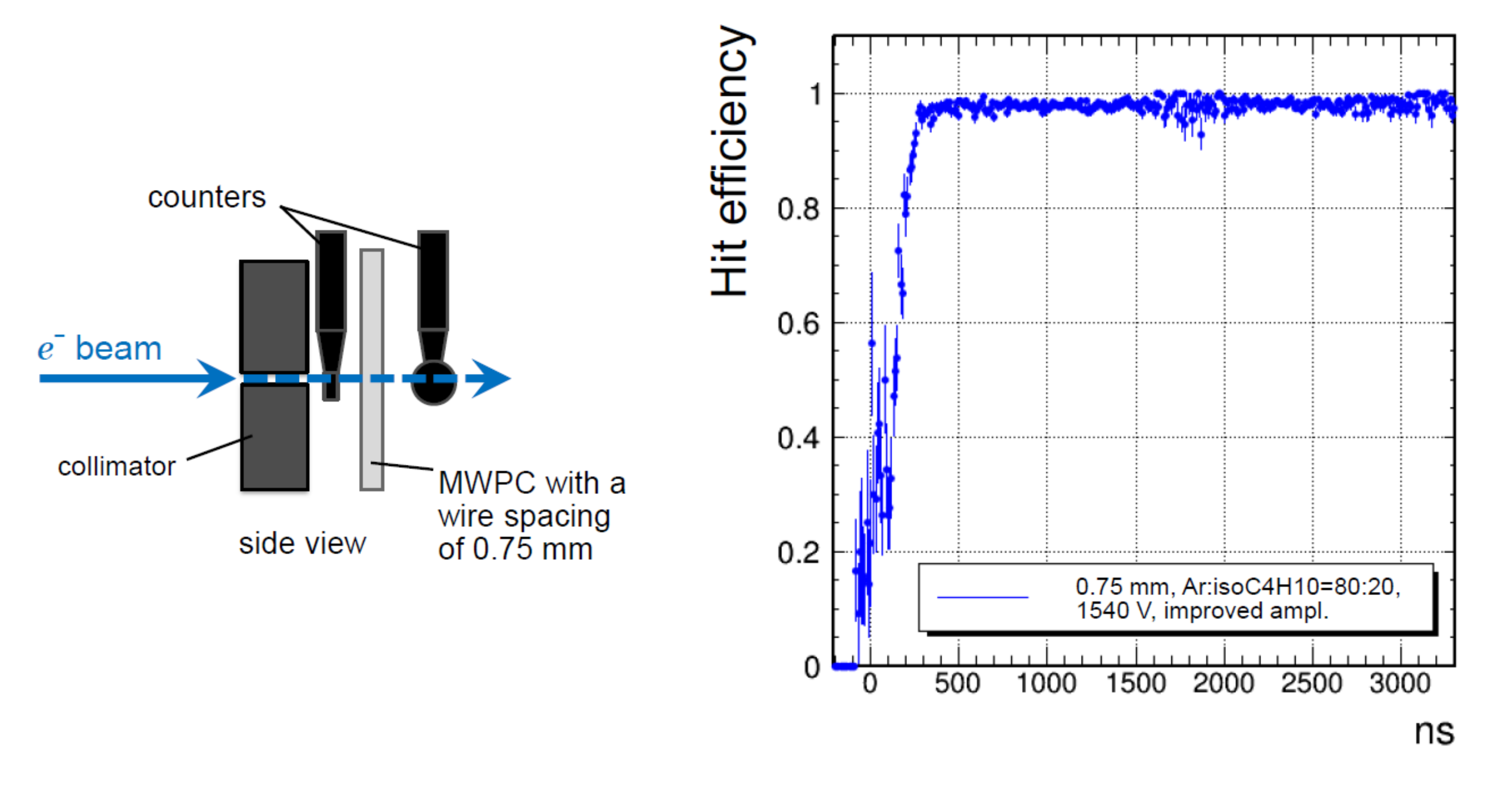}
        \caption{Setup (left) and hit efficiency of the MWPC (right) obtained from the test at Institute for Integrated Radiation and Nuclear Science, Kyoto University.}
        \label{fig:efficiafter}
      \end{minipage} &
      \begin{minipage}[t]{0.35\hsize}
        \centering
        \includegraphics[width=5.4cm,keepaspectratio,clip]{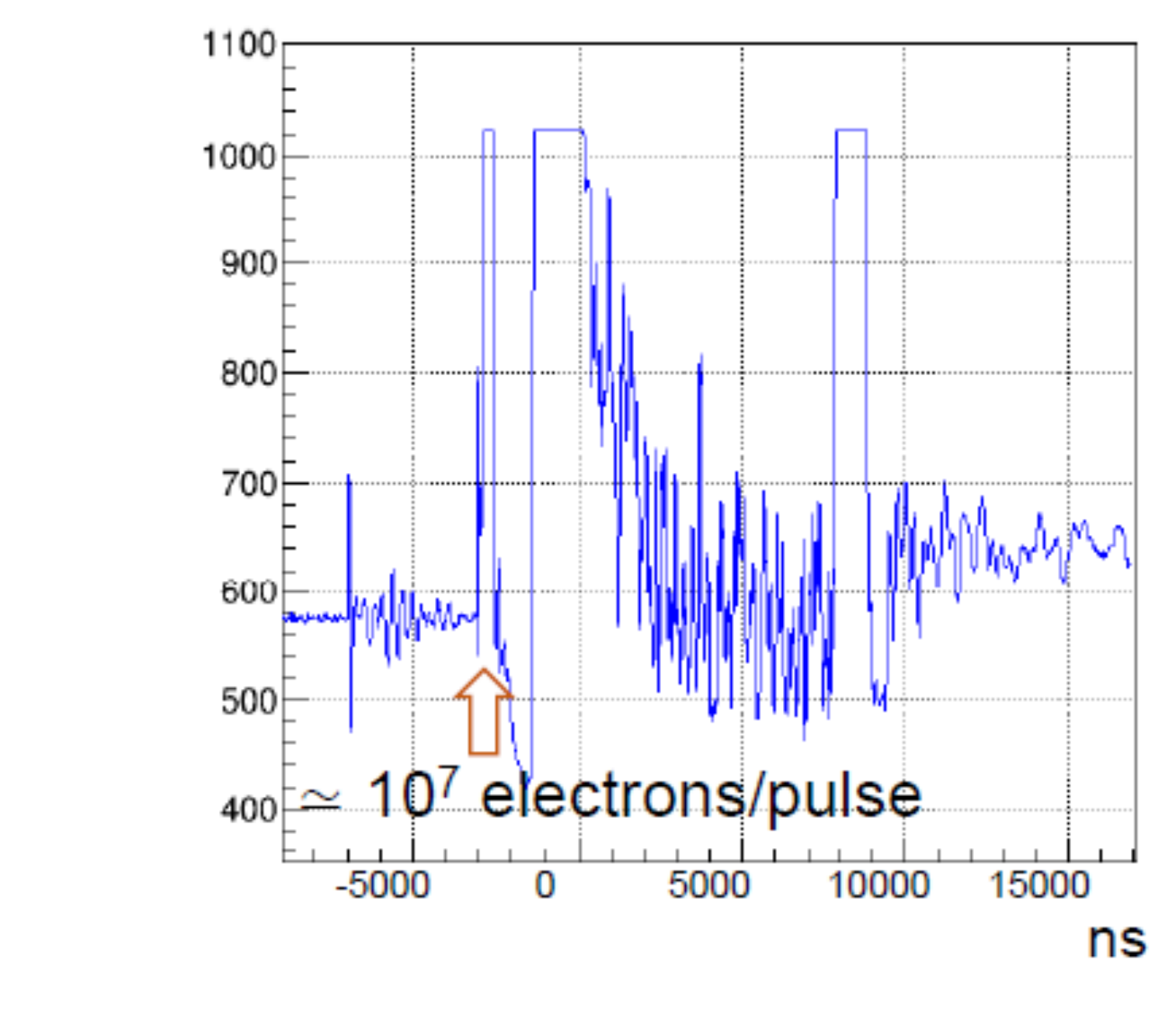}
        \caption{Waveform obtained when electron beams with the intensity equivalent to the prompt burst hit the MWPC.}
        \label{fig:promptburst}
      \end{minipage}
    \end{tabular}
  \end{figure}
\end{center}

We performed a beam test in February, 2018 at Institute for Integrated Radiation and Nuclear Science, Kyoto University. For single electron, hit efficiency was measured to be about 98\% at $300\ \mathrm{ns}$ after the MWPC starting to work (see Fig. \ref{fig:efficiafter}). But we observed random spikes in waveform when beams with the intensity equivalent to the prompt burst hit the MWPC as shown in Fig. \ref{fig:promptburst}. Electrons emitted from the cathode planes by ions might be a cause of these pulses. We have a plan to mix freon with the filling gas to absorb the electrons between the cathode and anode planes.
\vspace{-50pt}
\mbox{}\\
\section{Conclusion}
The DeeMe experiment aims to search for $\mu$-$e$ conversion with a single event sensitivity of $1\times10^{-13}$ for a graphite target down to $2\times10^{-14}$ for a silicon carbide target. The signal is an electron with a monochromatic energy of $105\ \mathrm{MeV}$, and we will search for it by using a magnetic spectrometer, which consists of a magnet and four MWPCs. \\
\indent
By optimizing the gas mixture filling the MWPCs and increasing the negative range of the readout amplifiers, hit-efficiency has been improved for single electrons. To absorb random-spike signals observed when the prompt burst hit the MWPC, the gas needs to be optimized a little more. We have a plan to mix freon with the filling gas.

\nolinenumbers

\end{document}